%% file: main.tex
\documentclass{ifacconf}
\usepackage{amsmath, amssymb}
\usepackage{graphicx}
\usepackage{cuted}
\usepackage{capt-of}
\usepackage{standalone}
\usepackage{nicematrix}
\usepackage{tikz}
\usetikzlibrary{matrix, calc, positioning, arrows, arrows.meta, fit, decorations.pathreplacing, shadings, backgrounds, shapes}

\usepackage{graphicx}
\usepackage{siunitx}

\usepackage{pgfplots}
\usepackage{contour}
\usepackage[dvipsnames]{xcolor}
\contourlength{0.7pt}
\pgfplotsset{compat=1.18}
\usepgfplotslibrary{groupplots}
\usetikzlibrary{arrows.meta}
\usepackage{booktabs}
\usepackage{ifthen}

\usepackage{natbib}

\begin{document}
\newcommand*{\vertbar}{\rule[-1ex]{0.5pt}{2.5ex}}
\newcommand*{\horzbar}{\rule[.5ex]{2.5ex}{0.5pt}}

\begin{frontmatter}

\title{Explainable LP-MPC: Shadow Price Contributions Reveal MV-CV Pairings}

\thanks[footnoteinfo]{Corresponding Author: Dan O'Connor (e-mail: doconnor@controlconsulting.com).}

\author[1,2]{Lim C. Siang},
\author[3]{Daniel L. O'Connor} 

\address[1]{Burnaby Refinery, Burnaby, BC, Canada}
\address[2]{Georgia Institute of Technology, Atlanta, GA, United States}
\address[3]{Control Consulting Inc., Cascade, MT, United States}

\begin{abstract}
In the process industries, MPC (Model Predictive Control) is typically implemented as a two-stage controller with a Linear Program (LP) steady-state optimizer that generates economically optimal targets for the MPC algorithm. Abnormal behaviors in industrial LP optimizers are often difficult to rationalize, especially when a large number of manipulated variables (MVs) and controlled variables (CVs) are involved. We introduce a novel, post-hoc LP explainability method by recasting the role of shadow prices in the LP solution as an attribution mechanism for MV-CV relationships. The core idea is that the shadow price of a constrained CV is not just an intrinsic property of the LP solution, but can be split into contributions from individual unconstrained MVs and resolved into one-to-one MV-CV pairings using a linear sum assignment algorithm. The proposed MV-CV pairing framework serves as a practical explainability tool for online LP-MPC systems, enabling practitioners to diagnose suboptimal constraints and verify alignment of the controller's behavior with its original design.
\end{abstract}


\end{frontmatter}

\begin{strip}
    \centering
    \includegraphics[width=0.99\textwidth]{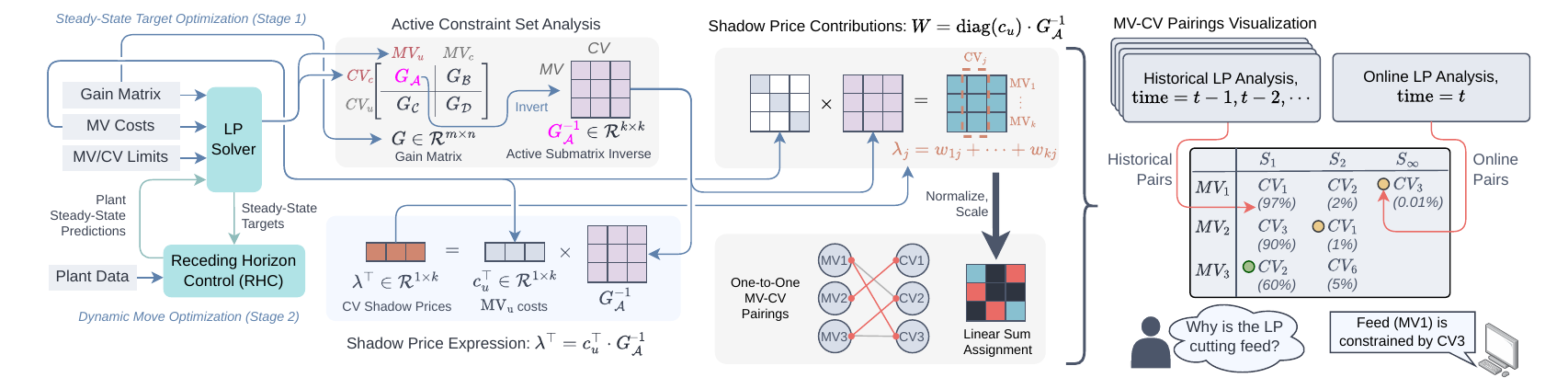}
    \captionof{figure}{The proposed MV-CV pairing framework enables practitioners to diagnose unexpected active constraint sets.}
    \label{fig:workflow}
\end{strip}

\section{Problem Statement}

In the chemical process industries, model predictive control (MPC) is commonly implemented in two separate stages, consisting of (1) a steady-state target optimization stage that calculates economically optimal targets, and (2) a receding horizon control (RHC) stage that moves the plant to these optimization targets. For clarity, the term `LP' in this paper refers exclusively to the steady-state target optimization stage implemented as a linear program (LP), which is widespread in the refining and petrochemical industries. This two-stage architecture has been given various names in the academic literature, including double-layer MPC, two-stage MPC, and LP-MPC cascade systems \citep{qin_survey_2003, Godoy2017, Elnawawi2022}.

In the steady-state LP optimization stage, limits on the manipulated variables (MVs) and controlled variables (CVs) are imposed in the form of linear inequality constraints, with economic costs defined on incremental MV moves. Some LP formulations may impose costs on CVs \citep{qin_survey_2003}, but these alternative formulations are not considered in this paper. The steady-state MV and CV targets are computed by solving an LP problem for cost minimization. The LP calculations are based on the steady-state gain matrix, the current steady-state operating conditions, and MV/CV limits \citep{maciejowski2002predictive}. At each program execution, the LP updates the steady-state targets to the RHC stage in response to changes in the current plant steady-state conditions (e.g. material composition changes and ambient conditions etc.), and operating limits (e.g. feed rate targets, product specifications etc.). 

On the overarching topic of plantwide control, self-optimizing control \citep{skogestad2000selfoptimizing} is an offline design methodology for CV selection such that constant-setpoint operation remains economically near-optimal under disturbances. Recent extensions systematically handle changing active constraint regions through advanced regulatory control (ARC) elements \citep{Skogestad2023AdvancedControl}. In contrast, the LP-MPC architecture identifies the active constraint region at run-time, and serves as a form of complex MV-CV switching to handle changing active constraint regions through the LP solution. This online formulation scales to industrial controllers with tens to hundreds of MVs and CVs, where the combinatorial complexity of possible active constraint sets may be intractable for systematic offline design of the switching regions. In practice, these industrial LP-MPC systems are generally $m$~CV $\times$ $n$~MV non-square controllers designed with more CVs than MVs ($m > n$) to honor all plant operating limits.

Due to the vertex-driven nature of the LP, even a single disturbance or a minor shift in operating limits could trigger a change in the active set of constraints in the LP solution, fundamentally altering the control strategy at every execution cycle. Despite the white-box nature of the LP model formulation, an important open problem for large-dimensional industrial LP-MPC controllers is addressing the difficulty around explaining the LP solution \citep{Elnawawi2022}. Systematic approaches for reasoning about real-world LP-MPC controllers remain elusive in practice, creating a disconnect between academic MPC developments and the practical needs of engineers and operators in industry. 

Even for a modest number of MVs and CVs, the combinatorial complexity of MV-CV interactions poses challenges for human understanding of the LP solution. In practice, plant personnel often need to review large, static tables to answer questions like ``\textit{why is the LP cutting feed rate?}”. Without a satisfactory answer to rationalize controller actions, operators may switch off the entire controller.  These issues represent a critical support gap after controller commissioning, and the need for post-hoc explainability tools to troubleshoot controller behavior in an intuitive and self-service manner \citep{Siang2022}.

\section{Methodology}
We introduce a novel, post-hoc analytical method to interrogate the steady-state target optimization LP solution and explain which unconstrained MV is primarily responsible for enforcing an active CV constraint. Our approach results in one-to-one MV-CV pairs that are physically interpretable by practitioners to understand controller behavior, as illustrated in Fig.~\ref{fig:workflow}.

We first recall a standard industrial LP-MPC formulation and the associated Karush-Kuhn-Tucker (KKT) optimality conditions. At the optimum, we restrict attention to the active constraint set and derive an expression for the CV shadow prices in terms of the inverse active gain matrix and economic costs on MVs. We show that the shadow price of a constrained CV can be expressed in an unreduced form that shows the contributions of each unconstrained MV, defining a shadow price contribution matrix.  After applying sign-correction and scaling on the matrix, one-to-one MV-CV pairings are identified using a linear sum assignment algorithm.

\subsection{General LP-MPC Optimizer Formulation}

A brief review of the general form of the LP-MPC problem is presented here. Details of industrial LP-MPC formulations are described elsewhere in the literature \citep{maciejowski2002predictive, qin_survey_2003, Nikandrov2009sensitivity}. Throughout, we assume that at each control interval, the LP solution is non-degenerate and feasible, with infeasibility handled by a CV ranking approach \citep{qin_survey_2003}. We interrogate the LP solution at a single control interval, thus, the LP formulation is time-independent. It is also customary to formulate the LP problem in terms of deviation variables or incremental moves ($\Delta MV$) to simplify engineering analysis \citep{rao1999steady, Godoy2017}.

Consider an LP-MPC controller with steady‐state gain matrix  \(G \in \mathbb{R}^{m \times n}\) relating \(m\) rows of CVs to \(n\) columns of MVs. At a particular control interval, the steady-state LP optimizer solves for an optimal set of incremental MV moves, $\Delta MV^* \in \mathbb{R}^{n \times 1}$, subject to upper ($U$) and lower ($L$) operating limits on the MVs and CVs.  The predicted changes in CV steady-state values, $\Delta CV^{ss*} \in \mathbb{R}^{m \times 1}$, are given by $\Delta CV^{ss*} = G \cdot \Delta MV^*$.  The allowable incremental MV changes based on the current MV values and their limits are $\Delta MV_L = MV_L - MV$ and $\Delta MV_U = MV_U - MV$.  The allowable incremental CV$^{ss}$ changes based on the current CV$^{ss}$ values and their limits are $\Delta CV^{ss}_L = CV_L -  CV^{\mathrm{ss}}$ and $\Delta CV^{ss}_U = CV_U - CV^{\mathrm{ss}}$. 

The overall time-independent LP problem at a single control interval has 4 inequality constraints:
\begin{equation}
\begin{aligned}
\min_{\Delta MV} \quad & c^{\mathsf T}\Delta MV \\
\mathrm{s.t.} \quad
& \Delta CV_L^{ss} - G\,\Delta MV \le 0, \\
& G\,\Delta MV - \Delta CV_U^{ss} \le 0, \\
& \Delta MV_L - \Delta MV \le 0, \\
& \Delta MV - \Delta MV_U \le 0.
\end{aligned}
\label{eq:mpc_lp}
\end{equation}
where \( c \in \mathbb{R}^{n \times 1} \) is the column vector that represents economic costs on $\Delta MV$. To analyze the LP optimality conditions, we combine the objective and inequality constraints into the Lagrangian, $\mathcal{L}$ with non-negative shadow prices $\lambda_L,\lambda_U, \mu_L,\mu_U \ge 0$ for each inequality ($\lambda$: CV, $\mu$: MV). Applying the KKT conditions for optimality, the stationarity condition states that the gradient of the Lagrangian with respect to the LP decision variables, $\Delta MV$, must vanish at the LP solution:
\begin{equation}
\frac{\partial \mathcal{L}}{\partial \Delta MV}
= c - G^{\mathsf T}\lambda_L + G^{\mathsf T}\lambda_U - \mu_L + \mu_U = 0.
\label{eq:kkt_stationarity}
\end{equation}
Commercial LP-MPC systems report the shadow price for each constrained variable instead of each inequality by collapsing the multipliers into a net shadow price term. Following this practice, we define the net, signed shadow prices as $\lambda \equiv \lambda_L - \lambda_U$ and $\mu \equiv \mu_L - \mu_U$. Algebraic manipulation of \eqref{eq:kkt_stationarity} yields the stationarity equation in terms of the net shadow prices:
\begin{equation}
c - G^{\mathsf T}\lambda - \mu = 0.
\label{eq:kkt_stationarity_net}
\end{equation}
The shadow prices $\lambda$ and $\mu$ represent the marginal change in the LP objective function if a constraint is relaxed by one unit to allow further incremental MV movement. 

\subsection{Active Constraint Set Analysis}

A critical point of understanding is that each active CV constraint bounded at its limit will consume a degree of freedom from the unconstrained MVs. Thus, at the LP solution, the number of unconstrained MVs is equal to the number of constrained CVs, forming a square \textit{active constraint set}, $\mathcal{A}$, which governs the control strategy. The gain matrix can be partitioned into 4 sub-matrices, $G_\mathcal{A}, G_\mathcal{B}, G_\mathcal{C}$ and $G_\mathcal{D}$, based on each variable's constraint.
\begin{equation}
G = 
\begin{bNiceArray}{w{c}{1.0cm}|w{c}{1.0cm}}[
    first-row,
    first-col,
    code-for-first-row = \scriptstyle,
    code-for-first-col = \scriptstyle,
    cell-space-limits = 3pt
]
 & \color{magenta}{MV_u} & MV_c \\
\color{magenta}{CV_c} & \color{magenta}{G_\mathcal{A}} & G_\mathcal{B} \\
\hline
CV_u & G_\mathcal{C} & G_\mathcal{D} \\
\end{bNiceArray}
\label{eq:partitionG}
\end{equation}
$MV_u$ represents unconstrained MVs and the degrees of freedom available to hold constrained CVs, $CV_c$ at their limits. $MV_c$ are constrained MVs that are saturated and not useful for controlling any CVs, and $CV_u$ are unconstrained CVs floating between limits. Knowing $G_\mathcal{A} \in \mathbb{R}^{k \times k}$ where $ k \equiv |CV_c| = |MV_u|$, allows us to impose further restrictions on the net stationarity equation in \eqref{eq:kkt_stationarity_net} by taking rows that only correspond to the unconstrained MVs for every term. By complementary slackness,  any unconstrained MV $i$ and unconstrained CV $j$ have zero shadow prices: 
$$\mu_i = 0 \quad \forall\, i \in MV_u, \qquad \lambda_j = 0 \quad \forall\, j \in CV_u$$
This analysis yields the restricted stationarity equation for the active constraint set: $c_u - G_\mathcal{A}^{\mathsf T} \lambda = 0 $ in column-vector form where \(c_{u} \;\in\; \mathbb{R}^{k\times1}\) is the vector of costs associated with $MV_u$ and \( \lambda \in \mathbb{R}^{k\times1} \) is the vector of shadow prices for $CV_c$. Tranposing this result using $(AB)^{\mathsf T} = B^{\mathsf T}A^{\mathsf T}$, we see equivalently in row-vector form $\lambda^{\mathsf T} G_\mathcal{A}= c_u^{\mathsf T}$. We assume the LP is feasible and non-degenerate, and that \( G_\mathcal{A} \) is square and invertible, allowing us to express the CV shadow prices analytically as a function of the inverse gain matrix:
\begin{equation}
\lambda^{\mathsf T} = c_u^{\mathsf T} G_\mathcal{A}^{-1}
\label{eq:lambda_expression}
\end{equation}
The inverse active gain submatrix, $G_\mathcal{A}^{-1}$, with MVs as rows and CVs as columns, uncovers any indirect relationships between MV and CVs that are not present in the open-loop gain matrix, $G$, due to the LP solution that's driving $MV_u$ to control $CV_c$ at their limits.

To our knowledge, the explicit mathematical expression between the shadow prices, costs and inverse square active gain matrix, has not appeared in the LP-MPC literature, even though the expression is algebraically immediate from the LP optimality conditions.

\subsection{Shadow Price Contributions by MVs}

From the matrix form of the CV shadow prices in \eqref{eq:lambda_expression}, it follows that for each CV $j$, we can express its shadow price as a sum of costs and inverse gains from its MVs:
\begin{equation}
\lambda_j = c_1\cdot {[G_\mathcal{A}^{-1}]}_{1,j} + c_2\cdot {[G_\mathcal{A}^{-1}]}_{2,j} + \cdots + c_k\cdot {[G_\mathcal{A}^{-1}]}_{k,j}
\label{eq:cv_shadow_price}
\end{equation}
Each term $w_{ij} = c_i \cdot {[G_\mathcal{A}^{-1}]}_{i,j} $ in \eqref{eq:cv_shadow_price} represents the contribution of $MV_i$ to the shadow price \( \lambda_j \) of $CV_j$.

To generalize \eqref{eq:cv_shadow_price} to all CVs, we define a \textit{shadow price contribution matrix} $W$ that describes, in unreduced form, the element-wise cost-weighted MV contributions before they are column-wise aggregated into the observed CV shadow prices from the LP solution in \eqref{eq:lambda_expression}:
\begin{equation}
W = \operatorname{diag}(c_u) \cdot G_\mathcal{A}^{-1}
\label{eq:raw_weight_matrix}
\end{equation}
The shadow price of each CV is recovered as a weighted sum along each CV column in the $W$ matrix, $\lambda_j = \sum_{i=1}^{k} w_{ij}$. $W$ explains the marginal cost of individual MV movements responsible for enforcing CV constraints in the active constraint set $\mathcal{A}$.

However, the physical interpretation of $W$ is complicated by the directional asymmetry and sign patterns on the CV shadow prices $\lambda$, as described earlier in \eqref{eq:kkt_stationarity_net}. If \(\lambda_j < 0\), the CV upper bound is constrained.

For a consistent interpretation in alignment with the cost-reducing direction of the LP objective, we apply a sign-adjusted correction to each column of the $W$ matrix:
\begin{equation}
W^{\mathrm{corr}} = W \cdot \operatorname{diag}(-\operatorname{sgn}(\lambda))
\label{eq:weighted_gain}
\end{equation}
After sign correction, \(W^{\mathrm{corr}}_{ij} < 0 \) represents a signed \textit{marginal improvement} in the LP objective, quantifying a cost‐minimizing contribution of $MV_i$ that is economically aligned with the LP objective, regardless of whether $CV_j$ is constrained at the upper or lower bound.

\subsection{Local MV-CV Pairing}

We can think of classical loop pairing as the `forward problem' for controller synthesis, where we ask which MVs and CVs should be paired together based on process physics and the active constraints we wish to control. In contrast, this LP-MPC pairing framework represents the `inverse problem' for controller diagnostics, where the control strategy has been decided by the LP optimizer based on both process physics and economics. We now want to reverse engineer the LP solution to recover the implicit MV-CV pairings. 

For each constrained CV, we define its \textit{locally most effective} MV pair as the choice that yields the greatest improvement in objective function per unit of CV relaxation. This corresponds to the MV row with the most negative entry for each CV column in $W^{\mathrm{corr}}$. However, this local pairing perspective does not account for competition across CVs for the same MV. Since multiple CVs may share the same MV as its locally most effective pair, we must further define a global pairing rule to resolve conflicts and produce a one-to-one MV–CV pairing assignment.

\subsection{Scale Invariance for Global MV-CV Pairing}

In practice, variables in the gain matrix typically span vastly different engineering units and numerical scales. \(W^{\mathrm{corr}}\) is scale-dependent with units of $\$/\Delta CV$, making a direct comparison of entries across multiple variables physically not meaningful without appropriate scaling \citep{Waller1995}. To ensure that all shadow price contributions have a common relative basis independent of engineering units, each CV column \(j\) is normalized by the absolute value of its most negative element in that column. This choice of CV scaling has several desirable properties: (1) it results in a scale-invariant, dimensionless shadow price contribution matrix \(\Pi_{ij}\), (2) each term represents normalized MV contributions to each CV's shadow price on a relative basis, (3) it preserves the CV's local ranking of preferred MV pairs in the most negative entry of $-1$ for each column.
\begin{equation}
\Pi_{ij} = \frac{W^{\mathrm{corr}}_{ij}}{\left| \min_p W^{\mathrm{corr}}_{pj}\right|}
\label{eq:normalized_weights}
\end{equation}
Scaling allows us to extend our local pairing definition to a global pairing rule. We define the set of one-to-one \textit{globally most effective} MV-CV pairings as the choice that minimizes the total normalized deviations from each CV’s locally most effective MV.

If every CV has a uniquely different local MV pair (i.e. the most negative entry for every CV columns all occur at different MV rows), the global MV-CV pairing rule simply reproduces the set of all local MV-CV pairings. When multiple CVs share the same locally most effective MV, each CV will be assigned to the next-best MV such that the aggregated, normalized deviations are minimized. 

A linear sum assignment algorithm is used to facilitate ``tie-breaking'' across multiple CVs in a consistent manner. To apply the linear sum assignment algorithm, an assignment penalty matrix \(P \in \mathbb{R}^{k \times k}\) encodes the normalized deviation associated with pairing $MV_i$ and $CV_j$:
\begin{equation}
P_{ij} =
\begin{cases}
\Pi_{ij}, & |\Pi_{ij}| \neq 0 \\
\infty, & \Pi_{ij} = 0
\end{cases}
\label{eq:penalty_matrix}
\end{equation}
The locally most effective MV for each CV is $P_{ij} = -1$, which is the smallest possible penalty in the matrix. For a non-existent CV-MV relationship in the inverse gain matrix (i.e. zeros), an infinite penalty of $P_{ij} = \infty$ is assigned to exclude these pairings from consideration. Small penalties \(P_{ij} < 0 \) encode deviations from the locally most effective MV and represent relative next-best choices for each CV. Larger penalties $P_{ij}>0$ are due to MV movements that are opposing the shadow price and moving away from the CV constraint, worsening the objective function. These are typically poor choices for pairing but will be selected if all other available MV choices are exhausted.

The optimal one-to-one assignment that minimizes the total assignment penalty is computed by solving a linear sum assignment problem:
\begin{align}
\min_{X \in \{0,1\}^{k \times k}} \quad & \sum_{i=1}^k \sum_{j=1}^k P_{ij} X_{ij} \label{eq:minpenalty} \\
\mathrm{s.t.} \quad 
& \sum_{i=1}^k X_{ij} = 1,\quad \sum_{j=1}^k X_{ij} = 1 \quad \forall i,j = 1,\dots,k, \nonumber
\end{align}

The assignment results are expressed using a binary matrix \(X \in \{0,1\}^{k \times k}\), where \(X_{ij} = 1\) if MV \(i\) is assigned to CV \(j\), and $0$ otherwise, which yields the globally most effective, minimum-penalty pairings of MVs and CVs.

\section{Results and Discussion}

\subsection{Physical Interpretation of MV-CV Pairings}

A trivial pairing arises when an MV affects only one CV, or vice-versa. The MV–CV pairing is effectively predetermined since there is only a single choice. In $P$, if there is a single, unique minima of \(-1\) across both row and column, the MV-CV pairing is also predetermined since any alternative choice would increase the total assignment cost and thus violate optimality in the linear sum assignment solution. These unambiguous pairings can be identified by iteratively removing these rows and columns in $P$ by inspection, even without using an assignment solver. The linear sum assignment algorithm resolves more economically unambiguous pairings, where each CV is assigned to its next-best MV such that the overall deviations from its preferred MV choices are minimized globally.

\subsection{Example: $2\times2$ Square Pairing Problem}

\newcommand{\GainA}{-0.115}   
\newcommand{\GainB}{0.001}   
\newcommand{\GainC}{3.090}    
\newcommand{\GainD}{0.080}   
\newcommand{\GainE}{-0.560}   
\newcommand{\GainF}{0.002}   

\newcommand{\CostOne}{12.5}    
\newcommand{\CostTwo}{0.1}   

\newcommand{\CVOneLower}{-0.7}
\newcommand{\CVOneUpper}{0.7}
\newcommand{\CVTwoLower}{-27}
\newcommand{\CVTwoUpper}{27}
\newcommand{\CVThreeLower}{-4}
\newcommand{\CVThreeUpper}{4}

Consider an LP-MPC system with gain matrix and LP costs defined as follows.
\[
\begin{aligned}
G =
\begin{bNiceArray}{w{c}{1.2cm}w{c}{1.2cm}}[
    first-row,
    first-col,
    code-for-first-row = \scriptstyle,
    code-for-first-col = \scriptstyle,
    cell-space-limits = 3pt
]
     & MV_1 & MV_2 \\
CV_1 & \GainA & \GainB \\
CV_2 & \GainC & \GainD \\
CV_3 & \GainE & \GainF \\
\end{bNiceArray}
\quad
c^{\mathsf T} =
  \begin{bmatrix}
  \CostOne  & \CostTwo
  \end{bmatrix}
\end{aligned}
\]

We assume that the $CV^{\mathrm{ss}}$ starting values are equidistant from the upper and lower CV limits.  The effective CV inequality constraints imposed are
\(
\CVOneLower \leq \Delta CV_1^{\mathrm{ss}} \leq \CVOneUpper,
\quad
\CVTwoLower \leq \Delta CV_2^{\mathrm{ss}} \leq \CVTwoUpper,
\quad
\CVThreeLower \leq \Delta CV_3^{\mathrm{ss}} \leq \CVThreeUpper,
\)
with no MV constraints. We can visualize the feasible region and geometry of the solution in terms of incremental MV movements as shown in the left subfigure in Fig.~\ref{fig:mv_cost_diagram}. The LP geometry shows that there are 4 possible solutions or vertices in the yellow feasible region, formed by CV constraints of $\Delta CV_1^{ss}$ and $\Delta CV_2^{ss}$. 

\begin{figure}[htbp]
    \centering
    \resizebox{0.99\columnwidth}{!}{%
        \input{fig_mv_cost_diagram.tex}%
    }
    \caption{Left -- Feasible region of LP solutions. Right -- Normal cones in cost space where each critical region $\mathcal{R}_i$ corresponds to an optimal vertex $V_i$.}
    \label{fig:mv_cost_diagram}
\end{figure}

The inequalities of $\Delta CV_3^{ss}$ shown in orange are outside of the feasible region and therefore $CV^{ss}_3$ will not participate in any optimum solution and our subsequent analysis reduces to a $2\times2$ sub-problem. The angle represented by the cost ratios, $r = (c_2/c_1)$, shown as a blue vector in Fig.~\ref{fig:mv_cost_diagram}, determines the direction of optimization and the optimal vertex in MV space. By solving the LP cost minimization problem as described in Eq. \eqref{eq:mpc_lp}, the optimum solution is $\Delta MV^* = [\pgfmathprintnumber[precision=1]{\SolX},\pgfmathprintnumber[precision=1]{\SolY}]$ found at vertex $V_2$. 
\subsubsection{MV-CV pairings for this LP solution}
The inverse matrix of the active constraint set and corresponding unconstrained MV costs are:
\[
\begin{aligned}
G_{\mathcal{A}}^{-1} =
\begin{bNiceArray}{w{c}{1.50cm}w{c}{1.50cm}}[
    first-row,
    first-col,
    code-for-first-row = \scriptstyle,
    code-for-first-col = \scriptstyle,
    cell-space-limits = 3pt
]
     & CV_1 & CV_2 \\
MV_1 & -6.5094 & 0.0814 \\
MV_2 & 251.4239 & 9.3572 \\
\end{bNiceArray}
\qquad
c_u^{\mathsf T}=\begin{bmatrix}\CostOne  & \CostTwo\end{bmatrix}
\end{aligned}
\]
By Eq. \eqref{eq:raw_weight_matrix} and Eq. \eqref{eq:normalized_weights}, adjusting for shadow price signs, $\lambda^{\mathsf T} = [-56.22, 1.95]$, we get:

\begin{minipage}{0.47\linewidth}
\[
W =
\begin{bNiceArray}{w{c}{0.90cm}w{c}{0.90cm}}[
    first-row,
    first-col,
    code-for-first-row = \scriptstyle,
    code-for-first-col = \scriptstyle,
    cell-space-limits = 3pt
]
     & CV_1 & CV_2 \\
MV_1 & -81.37 & 1.02 \\
MV_2 & 25.14  & 0.94 \\
\end{bNiceArray}
\]
\end{minipage}
\hfill
\begin{minipage}{0.47\linewidth}
\[
W^{\mathrm{corr}} =
\begin{bNiceArray}{w{c}{0.90cm}w{c}{0.90cm}}[
    first-row,
    first-col,
    code-for-first-row = \scriptstyle,
    code-for-first-col = \scriptstyle,
    cell-space-limits = 3pt
]
     & CV_1 & CV_2 \\
MV_1 & -81.37 & -1.02 \\
MV_2 & 25.14  & -0.94 \\
\end{bNiceArray}
\]
\end{minipage}

After scaling by the magnitude of the most negative entry in each column by Eq. \eqref{eq:normalized_weights}, we see that both CVs compete for MV1 as their locally most effective pair since they both have $-1$ entries in the first row:
\[
\Pi =
\begin{bNiceArray}{w{c}{0.95cm}w{c}{0.95cm}}[
    first-row,
    first-col,
    code-for-first-row = \scriptstyle,
    code-for-first-col = \scriptstyle,
    cell-space-limits = 3pt
]
     & CV_1 & CV_2 \\
MV_1 & -1 & -1 \\
MV_2 & 0.309 & -0.92 \\
\end{bNiceArray}
\]

There are only 2 possible pairing choices for this problem. 

\begin{itemize}
\item Diagonal: $\mathcal{S}_\mathrm{diag} = \{MV1\mapsto CV1,\,MV2\mapsto CV2\}$,
\item Off--diagonal: $\mathcal{S}_\mathrm{off} = \{MV1\mapsto CV2,\,MV2\mapsto CV1\}$.
\end{itemize}

The total penalties for these two assignments are $P_{\mathrm{diag}} = P_{11} + P_{22}$ and $P_{\mathrm{off}}  = P_{12} + P_{21}$. Since $P_{\mathrm{diag}} = -1.92 < P_{\mathrm{off}} = -0.69$, the diagonal pairings will minimize the overall penalty and the optimal set of assignments are $\mathcal{S}^*_\mathrm{diag} = \{MV1\mapsto CV1,\,MV2\mapsto CV2\}$.

\subsection{Sensitivity Analysis}

We perform a sensitivity analysis based on concepts from multiparametric linear programming (mp-LP) by sweeping the LP optimum solution as a function of cost ratios, $\Delta MV^*(r)$, as shown in the right subfigure in Fig.~\ref{fig:mv_cost_diagram}. Degenerate solutions arise when the cost level sets are parallel to an edge of the feasible region, and either vertex is optimal. Each critical region $\mathcal{R}_i$ represents the cone spanned by the normals of the active constraints at vertex $V_i$. Within these \textit{normal cones}, the optimal vertex $\Delta MV^*$ remains constant, with degenerate solutions at the cone boundaries. 

\subsubsection{Switching points for pairings}
Let $\Delta P = P_{\mathrm{diag}} - P_{\mathrm{off}}$. There exists a switching point for pairings when both penalties are equal with $\Delta P = 0$ because if $\Delta P < 0$, the diagonal pairing has a lower penalty, and vice-versa when $\Delta P > 0$. For this particular $2\times 2$ problem, by substituting the definitions and values for $P$ when both CVs are competing for the same MV, we see that:
\begin{equation}
\Delta P
= \frac{c_2}{|c_1|}
\left[
\frac{[G_\mathcal{A}^{-1}]_{21}\,\operatorname{sgn}(\lambda_1)}
     {|[G_\mathcal{A}^{-1}]_{11}|}
-
\frac{[G_\mathcal{A}^{-1}]_{22}\,\operatorname{sgn}(\lambda_2)}
     {|[G_\mathcal{A}^{-1}]_{12}|}
\right]
\label{eq:deltaP_factorized}
\end{equation}

For any cost ratios within the normal cone of the optimal vertex in Fig.~\ref{fig:mv_cost_diagram}, the shadow price sign patterns are fixed, so the bracketed term is constant, and the switching point \(\Delta P = 0\) reduces to $c_2 = 0$. If both CVs prefer MV2, by a similar algebraic exercise, $\Delta P = 0$ reduces to $c_1 = 0$. Note that the form of $\Delta P$ derived in \eqref{eq:deltaP_factorized} is only true when the local preferences for MVs are the same. In such cases, pairings will flip when the cost sign changes. This analysis shows that even for a simple $2\times 2$ problem, pairings are not solely determined by the LP optimum, which has important consequences: (1) the optimum is invariant to positive cost scalings if $r$ remains in the same normal cone, (2) the sign patterns of $c_1$ or $c_2$ may change the optimum if $r$ moves outside of the normal cone, (3) the pairings are influenced by the sign patterns of costs; when costs change, even when the optimum does not change, pairings may change.

\subsection{Visualization of MV-CV Pairings}

This MV-CV pairing framework is a practical solution for performing constraint analysis for both historical and online controller data. Constraint analysis is useful because when unexpected controller behavior occurs, plant personnel may turn off the entire LP-MPC controller or `clamp' MV limits (less often CV limits) to force the optimizer in certain directions that are operationally desired. This occurs for 3 primary reasons: (1) the objective of the controller is not well understood or the LP solution is counter-intuitive, (2) the economic objectives set by LP costs are incorrectly configured during controller design, or (3) gain consistency issues in the model cause the LP to incorrectly trade off constraints \citep{oconnor2022}.

\subsubsection{Visualization of historical pairings} First, we demonstrate how historical pairings can be visualized in practice. On an industrial controller (30 MVs, 63 CVs), we extracted the historical square active constraint sets, $\mathcal{A}_1, \mathcal{A}_2, \cdots, \mathcal{A}_N$ for an arbitrary 200-day window and evaluated the MV-CV pairings for all 1-minute control intervals. The truncated results (6 of 30 MVs) are presented in Table \ref{tab:mv_pairings} to illustrate key concepts without loss of generality. Each MV row contains three main columns $\mathcal{S}_1, \mathcal{S}_2, \mathcal{S}_3$ representing the top 3 historical CV pairings for that MV. The pairings are arranged in descending order across each row based on the percentage of time the MV-CV pair was found in historical data. For example, the first row informs us that MV2 is almost always unconstrained and paired with CV2 at its high limit, 99.4\% of the time. The fifth row informs us that MV17 is almost always constrained at its low limit and not paired with a CV, 96.5\% of the time.

\begin{table}[htbp]
\centering
\caption{Visualization of MV-CV Pairings}
\label{tab:mv_pairings}
\footnotesize
\renewcommand{\arraystretch}{1.0}

\newcommand{\gooddot}{\tikz[baseline=-0.6ex]{\fill[green!60!black] (0,0) circle (0.8ex);}}
\newcommand{\baddot}{\tikz[baseline=-0.6ex]{\fill[yellow!60!black] (0,0) circle (0.8ex);}}
\newcommand{\infeasibledot}{\tikz[baseline=-0.6ex]{\fill[red!60!black] (0,0) circle (0.8ex);}}
\newcommand{\topcell}[1]{\begin{tabular}[t]{@{}l@{}}#1\end{tabular}}

\begin{tabular}{@{} l p{1.5cm} p{1.5cm} p{1.5cm} | p{1.5cm} @{}}
\toprule
 & $\mathcal{S}_1$ & $\mathcal{S}_2$ & $\mathcal{S}_3$ & $\mathcal{S}_\infty$ \\ \midrule
MV2   & \topcell{\gooddot~CV2-HI \\ (\textit{99.4}\%)} 
      & \topcell{CV3-HI \\ (\textit{0.52}\%)} 
      & \topcell{CV9-HI \\ (\textit{0.05}\%)}
      & \topcell{} \\[1ex]

MV9   & \topcell{\gooddot~CV12-LO \\ (\textit{82.8}\%)} 
      & \topcell{MV-LO \\ (\textit{17.0}\%)} 
      & \topcell{OOS \\ (\textit{0.26}\%)}
      & \topcell{} \\[1ex]

MV13  & \topcell{CV14-HI \\ (\textit{96.4}\%)} 
      & \topcell{CV6-LO \\ (\textit{3.2}\%)} 
      & \topcell{MV-HI \\ (\textit{0.43}\%)} 
      & \topcell{\baddot~CV7-HI \\ (\textit{0.001}\%)} \\[1ex]

MV16  & \topcell{MV-LO \\ (\textit{55.5}\%)} 
      & \topcell{\gooddot~CV17-HI \\ (\textit{45.0}\%)}  
      & \topcell{CV14-HI \\ (\textit{0.006}\%)} 
      & \topcell{} \\[1ex]

MV17  & \topcell{\baddot~MV-LO \\ (\textit{96.5}\%)} 
      & \topcell{OOS \\ (\textit{3.5}\%)} 
      & \topcell{CV14-HI \\ (\textit{0.04}\%)}  
      & \topcell{} \\[1ex]

MV28  & \topcell{CV4-HI \\ (\textit{66.8}\%)} 
      & \topcell{\baddot~CV28-HI \\ (\textit{32.1}\%)} 
      & \topcell{CV27-HI \\ (\textit{0.87}\%)} 
      & \topcell{} \\ \midrule

Infeasible & \multicolumn{4}{l}{\infeasibledot~CV4, \infeasibledot~CV10} \\
\bottomrule
\end{tabular}

\end{table}

The organization of this table allows a large number of possible constraint combinations to be displayed in a compact form. The pairings follow a heavy-tailed distribution, a small number of pairings occur frequently, but a large number of pairings occur very infrequently. For this controller, 3 columns of pairings were enough to cover over $99$\% of the pairings observed for each of the 30 MVs. The last column $\mathcal{S}_\infty$ is used to highlight unusual pairings not represented by the top 3 pairings, and would likely be blank for most variables. 

\subsubsection{Ideal pairings}  This table-based constraint analysis explains how each degree of freedom, represented by each MV, was historically used by the LP-MPC controller. We emphasize that a clear engineering understanding of the controller's design objectives is needed to determine if a pairing is `ideal'. By comparing historical pairings against ‘ideal’ pairings, practitioners can verify alignment of the controller’s historical behaviour with its original design intent and control objectives. For instance, in a distillation column, bottoms steam flow rate is typically the primary MV for controlling the bottoms product purity CV. In the controller's original design intent, the `ideal' pairing is therefore steam MV with purity CV. 

If the table shows an MV with a significant fraction of operating time with non-ideal constraint pairings, further engineering investigation is warranted to understand if the controller is still functioning as intended. Examples of these investigative questions include, why is MV16 constrained at a low limit 55\% of the time? Why is MV13 paired with an unusual CV7? Why is MV17 out of service (OOS) 3.5\% of the time, or why is MV9 hitting a low MV limit 17\% of the time? By investigating the reasons behind those historical constraints, practitioners can estimate the economic benefits of constraint relief, and make adjustments to operating practices or justify future capital projects.

\subsubsection{Visualization of online pairings}  We assume `ideal' pairings are available and superimpose the real time pairings from an online controller as colored dots to the table.  This offers an effective visualization approach to analyze the current LP constraints in relation to historical precedence. Green dots indicate `ideal' pairings that match the controller's design intent. Yellow dots flag `non-deal' pairings for review, implying that the MV is incorrectly clamped or paired with an undesired CV. Red dots show lower-ranked CVs that were given up due to LP infeasibilities, which warrant further engineering investigations to understand the reasons why. We emphasize that a pairing is considered `ideal' if it aligns with the controller's design intent, even if it is not the most frequent pair found in $\mathcal{S}_1$. Conversely, a historically frequent pairing in $\mathcal{S}_1$ does not necessarily mean it is ideal. For industrial LP-MPC controllers, just because the controller is `ON' with a high uptime, it does not necessarily mean that it is running effectively.

Even if the original design intent is unclear, the historical constraints in the table provides a baseline reference for how the controller has typically behaved. For example in MV17, by reducing the MV17 low limit to provide constraint relief, it is possible to provide immediate feedback to operators showing how the MV17 degree of freedom is used in the current active constraint set. As another online example, if there is a pairing reassignment of the steam MV to another unusual CV such as the column's dP, it could indicate flooding or hydraulic limitations. By observing how pairings are reassigned in real-time, operators and engineers would immediately understand why the controller is cutting steam.

\section{Conclusions}
We present an explainable LP-MPC framework based on shadow prices for understanding which MV is primarily driving a CV's constraint, which represents a novel conceptual leap forward in the search for interpretable LP-MPC controller targets in the chemical process industries. Our methodology naturally admits multiple valid ways of normalization to construct a dimensionless contribution matrix. Empirically, by evaluating large real-world controllers up to $65~\text{MVs} \times 65~\text{CVs}$, we observed that column normalization based on the largest negative entry resulted in the most consistently `correct' pairings that made engineering sense. However, there is no universally accepted ground truth for these implicit LP-derived MV–CV pairings, which remains an open question for future research. Despite these conceptual limitations, the proposed framework offers practitioners a consistent and systematic way to interrogate LP solutions and make the optimizer's implicit decision-making logic more transparent. Taking a MV-CV pairing approach for constraint analysis provides valuable diagnostic insights for understanding current and historical controller behavior. Unexpected pairing mismatches can reveal potential controller issues and guide practitioners toward actionable opportunities for constraint relief.

\bibliography{ifacconf}             
                                                   
\end{document}

%% file: fig_mv_cost_diagram.tex
\begin{tikzpicture}

\providecommand{\GainA}{-0.11}   
\providecommand{\GainB}{0.001}   
\providecommand{\GainC}{3.090}   
\providecommand{\GainD}{0.080}   
\providecommand{\GainE}{-0.50}   
\providecommand{\GainF}{0.002}   

\pgfmathsetmacro{\gA}{\GainA}
\pgfmathsetmacro{\gB}{\GainB}
\pgfmathsetmacro{\gC}{\GainC}
\pgfmathsetmacro{\gD}{\GainD}
\pgfmathsetmacro{\gE}{\GainE}
\pgfmathsetmacro{\gF}{\GainF}

\providecommand{\CVOneLower}{-0.7}
\providecommand{\CVOneUpper}{ 0.7}
\providecommand{\CVTwoLower}{-27}
\providecommand{\CVTwoUpper}{ 27}
\providecommand{\CVThreeLower}{-4}
\providecommand{\CVThreeUpper}{ 4}

\pgfmathsetmacro{\CVoneL}{\CVOneLower}
\pgfmathsetmacro{\CVoneU}{\CVOneUpper}
\pgfmathsetmacro{\CVtwoL}{\CVTwoLower}
\pgfmathsetmacro{\CVtwoU}{\CVTwoUpper}
\pgfmathsetmacro{\CVthreeL}{\CVThreeLower}
\pgfmathsetmacro{\CVthreeU}{\CVThreeUpper}

\providecommand{\CostOne}{10}    
\providecommand{\CostTwo}{0.1}   

\pgfmathsetmacro{\cOneVec}{\CostOne}
\pgfmathsetmacro{\cTwoVec}{\CostTwo}

\pgfmathsetmacro{\figsize}{7cm}
\path (-1,-1) rectangle (12.2,2.4);

\pgfmathsetmacro{\xmin}{-10}
\pgfmathsetmacro{\xmax}{ 10}
\pgfmathsetmacro{\ymin}{-600}
\pgfmathsetmacro{\ymax}{ 600}

\pgfmathsetmacro{\dxAxis}{\xmax-\xmin}
\pgfmathsetmacro{\dyAxis}{\ymax-\ymin}

\pgfmathsetmacro{\cOneMin}{-40}
\pgfmathsetmacro{\cOneMax}{ 40}
\pgfmathsetmacro{\cTwoMin}{-1}
\pgfmathsetmacro{\cTwoMax}{ 1}
\pgfmathsetmacro{\dxC}{\cOneMax - \cOneMin}
\pgfmathsetmacro{\dyC}{\cTwoMax - \cTwoMin}

\pgfmathsetmacro{\cMag}{sqrt(\cOneVec*\cOneVec + \cTwoVec*\cTwoVec)}
\pgfmathsetmacro{\ucx}{\cOneVec / \cMag}
\pgfmathsetmacro{\ucy}{\cTwoVec / \cMag}


\pgfmathsetmacro{\detG}{\gA*\gD - \gB*\gC}

\pgfmathsetmacro{\mOne}{-(\gA)/(\gB)}
\pgfmathsetmacro{\bOneU}{\CVoneU/(\gB)}
\pgfmathsetmacro{\bOneL}{\CVoneL/(\gB)}

\pgfmathsetmacro{\mTwo}{-(\gC)/(\gD)}
\pgfmathsetmacro{\bTwoU}{\CVtwoU/(\gD)}
\pgfmathsetmacro{\bTwoL}{\CVtwoL/(\gD)}

\pgfmathsetmacro{\mThree}{-(\gE)/(\gF)}
\pgfmathsetmacro{\bThreeU}{\CVthreeU/(\gF)}
\pgfmathsetmacro{\bThreeL}{\CVthreeL/(\gF)}

\pgfmathsetmacro{\xLL}{(\CVoneL*\gD - \gB*\CVtwoL)/\detG}
\pgfmathsetmacro{\yLL}{(\gA*\CVtwoL - \CVoneL*\gC)/\detG}

\pgfmathsetmacro{\xUL}{(\CVoneU*\gD - \gB*\CVtwoL)/\detG}
\pgfmathsetmacro{\yUL}{(\gA*\CVtwoL - \CVoneU*\gC)/\detG}

\pgfmathsetmacro{\xUU}{(\CVoneU*\gD - \gB*\CVtwoU)/\detG}
\pgfmathsetmacro{\yUU}{(\gA*\CVtwoU - \CVoneU*\gC)/\detG}

\pgfmathsetmacro{\xLU}{(\CVoneL*\gD - \gB*\CVtwoU)/\detG}
\pgfmathsetmacro{\yLU}{(\gA*\CVtwoU - \CVoneL*\gC)/\detG}

\pgfmathparse{\detG > 0 ? 1 : 0}
\let\orientation\pgfmathresult

\ifnum\orientation=1
  \def\vOneX{\xLL} \def\vOneY{\yLL}
  \def\vTwoX{\xUL} \def\vTwoY{\yUL}
  \def\vTriX{\xUU} \def\vTriY{\yUU}
  \def\vForX{\xLU} \def\vForY{\yLU}
\else
  \def\vOneX{\xLL} \def\vOneY{\yLL}
  \def\vTwoX{\xLU} \def\vTwoY{\yLU}
  \def\vTriX{\xUU} \def\vTriY{\yUU}
  \def\vForX{\xUL} \def\vForY{\yUL}
\fi

\def\VLabelOne  {$V_1$}
\def\VLabelTwo  {$V_2$}
\def\VLabelTri  {$V_3$}
\def\VLabelFor  {$V_4$}

\def\ELabelTwelve      {\scriptsize $E_{3,4}$}
\def\ELabelTwentyThree {\scriptsize $E_{4,1}$}
\def\ELabelThirtyFour  {\scriptsize $E_{1,2}$}
\def\ELabelFortyOne    {\scriptsize $E_{2,3}$}

\def\RLabelDA {$\mathcal{R}_1$}
\def\RLabelAB {$\mathcal{R}_2$}
\def\RLabelBC {$\mathcal{R}_3$}
\def\RLabelCD {$\mathcal{R}_4$}

\pgfmathsetmacro{\cx}{(\vOneX + \vTwoX + \vTriX + \vForX)/4}
\pgfmathsetmacro{\cy}{(\vOneY + \vTwoY + \vTriY + \vForY)/4}

\pgfmathsetmacro{\labelscale}{0.08}
\pgfmathsetmacro{\dx}{(\xmax-\xmin)*\labelscale}
\pgfmathsetmacro{\dy}{(\ymax-\ymin)*\labelscale}

\pgfmathsetmacro{\sxOne}{(\vOneX-\cx)>=0 ? 1 : -1} \pgfmathsetmacro{\syOne}{(\vOneY-\cy)>=0 ? 1 : -1}
\pgfmathsetmacro{\sxTwo}{(\vTwoX-\cx)>=0 ? 1 : -1} \pgfmathsetmacro{\syTwo}{(\vTwoY-\cy)>=0 ? 1 : -1}
\pgfmathsetmacro{\sxTri}{(\vTriX-\cx)>=0 ? 1 : -1} \pgfmathsetmacro{\syTri}{(\vTriY-\cy)>=0 ? 1 : -1}
\pgfmathsetmacro{\sxFor}{(\vForX-\cx)>=0 ? 1 : -1} \pgfmathsetmacro{\syFor}{(\vForY-\cy)>=0 ? 1 : -1}

\pgfmathsetmacro{\lxOne}{\vOneX + \sxOne*\dx} \pgfmathsetmacro{\lyOne}{\vOneY + \syOne*\dy}
\pgfmathsetmacro{\lxTwo}{\vTwoX + \sxTwo*\dx} \pgfmathsetmacro{\lyTwo}{\vTwoY + \syTwo*\dy}
\pgfmathsetmacro{\lxTri}{\vTriX + \sxTri*\dx} \pgfmathsetmacro{\lyTri}{\vTriY + \syTri*\dy}
\pgfmathsetmacro{\lxFor}{\vForX + \sxFor*\dx} \pgfmathsetmacro{\lyFor}{\vForY + \syFor*\dy}
\xdef\SolX{\vForX} \xdef\SolY{\vForY}

\pgfmathsetmacro{\edgelabelscale}{0.13}
\pgfmathsetmacro{\edx}{(\xmax-\xmin)*\edgelabelscale}
\pgfmathsetmacro{\edy}{(\ymax-\ymin)*\edgelabelscale}

\pgfmathsetmacro{\mTwelveX}{0.5*(\vOneX+\vTwoX)}
\pgfmathsetmacro{\mTwelveY}{0.5*(\vOneY+\vTwoY)}
\pgfmathsetmacro{\eTwelveDx}{\vTwoX-\vOneX}
\pgfmathsetmacro{\eTwelveDy}{\vTwoY-\vOneY}
\pgfmathsetmacro{\nTwelveX}{\eTwelveDy/\dyAxis}
\pgfmathsetmacro{\nTwelveY}{-\eTwelveDx/\dxAxis}
\pgfmathsetmacro{\dotTwelve}{\nTwelveX*(\mTwelveX-\cx)+\nTwelveY*(\mTwelveY-\cy)}
\pgfmathsetmacro{\signTwelve}{\dotTwelve>=0 ? 1 : -1}
\pgfmathsetmacro{\nOutTwelveX}{\signTwelve*\nTwelveX}
\pgfmathsetmacro{\nOutTwelveY}{\signTwelve*\nTwelveY}
\pgfmathsetmacro{\normTwelve}{abs(\nOutTwelveX)+abs(\nOutTwelveY)}
\pgfmathsetmacro{\normTwelveSafe}{\normTwelve==0 ? 1 : \normTwelve}
\pgfmathsetmacro{\lTwelveX}{\mTwelveX + (\nOutTwelveX/\normTwelveSafe)*\edx}
\pgfmathsetmacro{\lTwelveY}{\mTwelveY + (\nOutTwelveY/\normTwelveSafe)*\edy}
\pgfmathsetmacro{\rOutTwelveX}{\signTwelve * \eTwelveDy}
\pgfmathsetmacro{\rOutTwelveY}{\signTwelve * (-\eTwelveDx)}

\pgfmathsetmacro{\mTwentyThreeX}{0.5*(\vTwoX+\vTriX)}
\pgfmathsetmacro{\mTwentyThreeY}{0.5*(\vTwoY+\vTriY)}
\pgfmathsetmacro{\eTwentyThreeDx}{\vTriX-\vTwoX}
\pgfmathsetmacro{\eTwentyThreeDy}{\vTriY-\vTwoY}
\pgfmathsetmacro{\nTwentyThreeX}{\eTwentyThreeDy/\dyAxis}
\pgfmathsetmacro{\nTwentyThreeY}{-\eTwentyThreeDx/\dxAxis}
\pgfmathsetmacro{\dotTwentyThree}{\nTwentyThreeX*(\mTwentyThreeX-\cx)+\nTwentyThreeY*(\mTwentyThreeY-\cy)}
\pgfmathsetmacro{\signTwentyThree}{\dotTwentyThree>=0 ? 1 : -1}
\pgfmathsetmacro{\nOutTwentyThreeX}{\signTwentyThree*\nTwentyThreeX}
\pgfmathsetmacro{\nOutTwentyThreeY}{\signTwentyThree*\nTwentyThreeY}
\pgfmathsetmacro{\normTwentyThree}{abs(\nOutTwentyThreeX)+abs(\nOutTwentyThreeY)}
\pgfmathsetmacro{\normTwentyThreeSafe}{\normTwentyThree==0 ? 1 : \normTwentyThree}
\pgfmathsetmacro{\lTwentyThreeX}{\mTwentyThreeX + (\nOutTwentyThreeX/\normTwentyThreeSafe)*\edx}
\pgfmathsetmacro{\lTwentyThreeY}{\mTwentyThreeY + (\nOutTwentyThreeY/\normTwentyThreeSafe)*\edy}
\pgfmathsetmacro{\rOutTwentyThreeX}{\signTwentyThree * \eTwentyThreeDy}
\pgfmathsetmacro{\rOutTwentyThreeY}{\signTwentyThree * (-\eTwentyThreeDx)}

\pgfmathsetmacro{\mThirtyFourX}{0.5*(\vTriX+\vForX)}
\pgfmathsetmacro{\mThirtyFourY}{0.5*(\vTriY+\vForY)}
\pgfmathsetmacro{\eThirtyFourDx}{\vForX-\vTriX}
\pgfmathsetmacro{\eThirtyFourDy}{\vForY-\vTriY}
\pgfmathsetmacro{\nThirtyFourX}{\eThirtyFourDy/\dyAxis}
\pgfmathsetmacro{\nThirtyFourY}{-\eThirtyFourDx/\dxAxis}
\pgfmathsetmacro{\dotThirtyFour}{\nThirtyFourX*(\mThirtyFourX-\cx)+\nThirtyFourY*(\mThirtyFourY-\cy)}
\pgfmathsetmacro{\signThirtyFour}{\dotThirtyFour>=0 ? 1 : -1}
\pgfmathsetmacro{\nOutThirtyFourX}{\signThirtyFour*\nThirtyFourX}
\pgfmathsetmacro{\nOutThirtyFourY}{\signThirtyFour*\nThirtyFourY}
\pgfmathsetmacro{\normThirtyFour}{abs(\nOutThirtyFourX)+abs(\nOutThirtyFourY)}
\pgfmathsetmacro{\normThirtyFourSafe}{\normThirtyFour==0 ? 1 : \normThirtyFour}
\pgfmathsetmacro{\lThirtyFourX}{\mThirtyFourX + (\nOutThirtyFourX/\normThirtyFourSafe)*\edx}
\pgfmathsetmacro{\lThirtyFourY}{\mThirtyFourY + (\nOutThirtyFourY/\normThirtyFourSafe)*\edy}
\pgfmathsetmacro{\rOutThirtyFourX}{\signThirtyFour * \eThirtyFourDy}
\pgfmathsetmacro{\rOutThirtyFourY}{\signThirtyFour * (-\eThirtyFourDx)}

\pgfmathsetmacro{\mFortyOneX}{0.5*(\vForX+\vOneX)}
\pgfmathsetmacro{\mFortyOneY}{0.5*(\vForY+\vOneY)}
\pgfmathsetmacro{\eFortyOneDx}{\vOneX-\vForX}
\pgfmathsetmacro{\eFortyOneDy}{\vOneY-\vForY}
\pgfmathsetmacro{\nFortyOneX}{\eFortyOneDy/\dyAxis}
\pgfmathsetmacro{\nFortyOneY}{-\eFortyOneDx/\dxAxis}
\pgfmathsetmacro{\dotFortyOne}{\nFortyOneX*(\mFortyOneX-\cx)+\nFortyOneY*(\mFortyOneY-\cy)}
\pgfmathsetmacro{\signFortyOne}{\dotFortyOne>=0 ? 1 : -1}
\pgfmathsetmacro{\nOutFortyOneX}{\signFortyOne*\nFortyOneX}
\pgfmathsetmacro{\nOutFortyOneY}{\signFortyOne*\nFortyOneY}
\pgfmathsetmacro{\normFortyOne}{abs(\nOutFortyOneX)+abs(\nOutFortyOneY)}
\pgfmathsetmacro{\normFortyOneSafe}{\normFortyOne==0 ? 1 : \normFortyOne}
\pgfmathsetmacro{\lFortyOneX}{\mFortyOneX + (\nOutFortyOneX/\normFortyOneSafe)*\edx}
\pgfmathsetmacro{\lFortyOneY}{\mFortyOneY + (\nOutFortyOneY/\normFortyOneSafe)*\edy}
\pgfmathsetmacro{\rOutFortyOneX}{\signFortyOne * \eFortyOneDy}
\pgfmathsetmacro{\rOutFortyOneY}{\signFortyOne * (-\eFortyOneDx)}

\pgfmathsetmacro{\normScale}{0.001} 

\pgfmathsetmacro{\uDAxRaw}{\rOutFortyOneX * \normScale}
\pgfmathsetmacro{\uDAyRaw}{\rOutFortyOneY * \normScale}

\pgfmathsetmacro{\uABxRaw}{\rOutTwelveX * \normScale}
\pgfmathsetmacro{\uAByRaw}{\rOutTwelveY * \normScale}

\pgfmathsetmacro{\uBCxRaw}{\rOutTwentyThreeX * \normScale}
\pgfmathsetmacro{\uBCyRaw}{\rOutTwentyThreeY * \normScale}

\pgfmathsetmacro{\uCDxRaw}{\rOutThirtyFourX * \normScale}
\pgfmathsetmacro{\uCDyRaw}{\rOutThirtyFourY * \normScale}

\pgfmathsetmacro{\uAmag}{sqrt(\uABxRaw*\uABxRaw + \uAByRaw*\uAByRaw)}
\pgfmathsetmacro{\uBmag}{sqrt(\uBCxRaw*\uBCxRaw + \uBCyRaw*\uBCyRaw)}
\pgfmathsetmacro{\uCmag}{sqrt(\uCDxRaw*\uCDxRaw + \uCDyRaw*\uCDyRaw)}
\pgfmathsetmacro{\uDmag}{sqrt(\uDAxRaw*\uDAxRaw + \uDAyRaw*\uDAyRaw)}

\pgfmathsetmacro{\uAx}{\uABxRaw/\uAmag} \pgfmathsetmacro{\uAy}{\uAByRaw/\uAmag}
\pgfmathsetmacro{\uBx}{\uBCxRaw/\uBmag} \pgfmathsetmacro{\uBy}{\uBCyRaw/\uBmag}
\pgfmathsetmacro{\uCx}{\uCDxRaw/\uCmag} \pgfmathsetmacro{\uCy}{\uCDyRaw/\uCmag}
\pgfmathsetmacro{\uDx}{\uDAxRaw/\uDmag} \pgfmathsetmacro{\uDy}{\uDAyRaw/\uDmag}

\pgfmathsetmacro{\costRadius}{sqrt(\dxC*\dxC + \dyC*\dyC)}
\pgfmathsetmacro{\ShadeLen}{1.2 * \costRadius}
\pgfmathsetmacro{\RayLenCost}{0.15 * \costRadius}
\pgfmathsetmacro{\RayLen}{1.00 * \costRadius}
\pgfmathsetmacro{\edgelabelextra}{1.85}

\newcommand{\plotSafeRay}[2]{%
    \pgfmathsetmacro{\nx}{#1}%
    \pgfmathsetmacro{\ny}{#2}%
    \pgfmathsetmacro{\mag}{sqrt(\nx*\nx + \ny*\ny)}%
    \pgfmathsetmacro{\ux}{\nx/\mag}%
    \pgfmathsetmacro{\uy}{\ny/\mag}%
    \addplot[magenta!80, thick] coordinates {(0,0) (\RayLen*\ux,\RayLen*\uy)};
}

\tikzset{
    vertexlabel/.style={
        fill=yellow,
        draw=black,
        inner sep=1.2pt,
        rounded corners=2pt,
        text=black,
        on layer=foreground
    },
    edgelabel/.style={
        fill=magenta!20,
        draw=magenta!70!black,
        inner sep=1.2pt,
        rounded corners=2pt,
        text=black,
        on layer=foreground
    },
    edgearrow/.style={
        -{Stealth[length=3mm,width=2mm]},
        very thick,
        magenta,
        on layer=foreground
    }
}

\begin{axis}[
    name=mvaxis,
    width=\figsize,
    height=\figsize,
    xlabel={$\Delta MV_1$},
    ylabel={$\Delta MV_2$},
    xmin=\xmin, xmax=\xmax,
    ymin=\ymin, ymax=\ymax,
    grid=both,
    major grid style={line width=0.2pt, draw=gray!20},
    minor grid style={line width=0.1pt, draw=gray!10},
    minor tick num=1,
    y tick label style={xshift=1pt},
    ylabel style={yshift=-20pt}
]

\addplot[black!50, very thin] coordinates {(0,\ymin) (0,\ymax)};
\addplot[black!50, very thin] coordinates {(\xmin,0) (\xmax,0)};
\addplot[mark=+,only marks,mark size=4pt,line width=0.6pt] coordinates {(0,0)};

\addplot[only marks,mark=*] coordinates {
  (\vOneX,\vOneY)
  (\vTwoX,\vTwoY)
  (\vTriX,\vTriY)
  (\vForX,\vForY)
};

\draw[edgearrow] (axis cs:\mTwelveX,\mTwelveY) -- (axis cs:\lTwelveX,\lTwelveY);
\node[edgelabel] at (axis cs:{\mTwelveX + (\nOutTwelveX/\normTwelveSafe)*\edx*\edgelabelextra},
                             {\mTwelveY + (\nOutTwelveY/\normTwelveSafe)*\edy*\edgelabelextra}) {\ELabelTwelve};
\draw[edgearrow] (axis cs:\mTwentyThreeX,\mTwentyThreeY) -- (axis cs:\lTwentyThreeX,\lTwentyThreeY);
\node[edgelabel] at (axis cs:{\mTwentyThreeX + (\nOutTwentyThreeX/\normTwentyThreeSafe)*\edx*\edgelabelextra},
                             {\mTwentyThreeY + (\nOutTwentyThreeY/\normTwentyThreeSafe)*\edy*\edgelabelextra}) {\ELabelTwentyThree};
\draw[edgearrow] (axis cs:\mThirtyFourX,\mThirtyFourY) -- (axis cs:\lThirtyFourX,\lThirtyFourY);
\node[edgelabel] at (axis cs:{\mThirtyFourX + (\nOutThirtyFourX/\normThirtyFourSafe)*\edx*\edgelabelextra},
                             {\mThirtyFourY + (\nOutThirtyFourY/\normThirtyFourSafe)*\edy*\edgelabelextra}) {\ELabelThirtyFour};
\draw[edgearrow] (axis cs:\mFortyOneX,\mFortyOneY) -- (axis cs:\lFortyOneX,\lFortyOneY);
\node[edgelabel] at (axis cs:{\mFortyOneX + (\nOutFortyOneX/\normFortyOneSafe)*\edx*\edgelabelextra},
                             {\mFortyOneY + (\nOutFortyOneY/\normFortyOneSafe)*\edy*\edgelabelextra}) {\ELabelFortyOne};

\addplot[black, thick, dash dot, domain=\xmin:\xmax, samples=2] {\mOne * x + \bOneU}
    node[sloped, below, pos=0.15, font=\tiny, text=black!70] 
    {$\mathrm{CV1:} \pgfmathprintnumber[fixed, precision=3]{\gA}x + 
      \pgfmathprintnumber[fixed, precision=3]{\gB}y \leq 
      \pgfmathprintnumber[fixed, precision=3]{\CVoneU}$};

\addplot[black, thick, dashed, domain=\xmin:\xmax, samples=2] {\mOne * x + \bOneL}
    node[sloped, above, pos=0.85, font=\tiny, text=black!70] 
    {$\mathrm{CV1:} \pgfmathprintnumber[fixed, precision=3]{\gA}x + 
      \pgfmathprintnumber[fixed, precision=3]{\gB}y \geq 
      \pgfmathprintnumber[fixed, precision=3]{\CVoneL}$};

\addplot[black, thick, dash dot, domain=\xmin:\xmax, samples=2] {\mTwo * x + \bTwoU}
    node[sloped, below, pos=0.4, font=\tiny, text=black!70] 
    {$\mathrm{CV2:} \pgfmathprintnumber[fixed, precision=3]{\gC}x + 
      \pgfmathprintnumber[fixed, precision=3]{\gD}y \leq 
      \pgfmathprintnumber[fixed, precision=3]{\CVtwoU}$};
    
\addplot[black, thick, dashed, domain=\xmin:\xmax, samples=2] {\mTwo * x + \bTwoL}
    node[sloped, above, pos=0.5, font=\tiny, text=black!70] 
    {$\mathrm{CV2:} \pgfmathprintnumber[fixed, precision=3]{\gC}x + 
      \pgfmathprintnumber[fixed, precision=3]{\gD}y \geq 
      \pgfmathprintnumber[fixed, precision=3]{\CVtwoL}$};    

\addplot[orange, thick, dash dot, domain=\xmin:\xmax, samples=2] {\mThree * x + \bThreeU};
\addplot[orange, thick, dash dot, domain=\xmin:\xmax, samples=2] {\mThree * x + \bThreeL};

\addplot[
    fill=yellow,
    fill opacity=0.10,
    line width=3.0pt,        
    draw=magenta,          
    draw opacity=0.25
] coordinates {
  (\vOneX,\vOneY)
  (\vTwoX,\vTwoY)
  (\vTriX,\vTriY)
  (\vForX,\vForY)
} -- cycle;
    
\node[vertexlabel] at (axis cs:\lxOne,\lyOne) {\VLabelTri};
\node[vertexlabel] at (axis cs:\lxTwo,\lyTwo) {\VLabelFor};
\node[vertexlabel] at (axis cs:\lxTri,\lyTri) {\VLabelOne};
\node[vertexlabel] at (axis cs:\lxFor,\lyFor) {\VLabelTwo};

\pgfmathsetmacro{\sCost}{(\cTwoVec/\cOneVec) * (\dxC/\dyC)}
\pgfmathsetmacro{\sMV}{\sCost * (\dyAxis/\dxAxis)}
\pgfmathsetmacro{\mvDx}{-1}
\pgfmathsetmacro{\mvDy}{\sMV * \mvDx}
\pgfmathsetmacro{\mvNorm}{sqrt(\mvDx*\mvDx + \mvDy*\mvDy)}
\pgfmathsetmacro{\mvDxN}{\mvDx / \mvNorm}
\pgfmathsetmacro{\mvDyN}{\mvDy / \mvNorm}
\pgfmathsetmacro{\mvScale}{55} 

\pgfmathsetmacro{\mvPlotX}{\mvScale * \mvDxN}
\pgfmathsetmacro{\mvPlotY}{\mvScale * \mvDyN}

\addplot[blue, ultra thick, -{Latex[length=2mm,width=2mm]}] 
    coordinates {(0,0) (\mvPlotX,\mvPlotY)}
    node[pos=1, anchor=south west, font=\normalsize, xshift=-3.5mm, yshift=-3.5mm] {$r$};

\end{axis}


\pgfmathsetmacro{\uEOneTwoX}{\uCx}
\pgfmathsetmacro{\uEOneTwoY}{\uCy}
\pgfmathsetmacro{\uETwoThreeX}{\uDx}
\pgfmathsetmacro{\uETwoThreeY}{\uDy}
\pgfmathsetmacro{\uEThreeFourX}{\uAx}
\pgfmathsetmacro{\uEThreeFourY}{\uAy}
\pgfmathsetmacro{\uEFourOneX}{\uBx}
\pgfmathsetmacro{\uEFourOneY}{\uBy}

\begin{axis}[
    name=costaxis,
    at={(mvaxis.south east)},
    anchor=south west,
    xshift=1.0cm,
    width=\figsize,
    height=\figsize,
    xlabel={$c_1$},
    ylabel={$c_2$},
    xmin=\cOneMin, xmax=\cOneMax,
    ymin=\cTwoMin, ymax=\cTwoMax,
    x dir=reverse,
    y dir=reverse,
    scaled y ticks=false,
    y tick label style={/pgf/number format/fixed},
    axis line style={black},
    ticks=both,
    grid=both,
    major grid style={line width=0.2pt, draw=gray!20},
    minor grid style={line width=0.1pt, draw=gray!10},
    minor tick num=1,
    y tick label style={xshift=1pt},
    ylabel style={yshift=-15pt}    
]

\begin{scope}
  \clip (axis cs:0,0) -- (axis cs:{\ShadeLen*\uAx},{\ShadeLen*\uAy})
        -- (axis cs:{\ShadeLen*\uBx},{\ShadeLen*\uBy}) -- cycle;
  \fill[blue!35, opacity=0.4] (axis cs:\cOneMin,\cTwoMin) rectangle (axis cs:\cOneMax,\cTwoMax);
\end{scope}

\begin{scope}
  \clip (axis cs:0,0) -- (axis cs:{\ShadeLen*\uBx},{\ShadeLen*\uBy})
        -- (axis cs:{\ShadeLen*\uCx},{\ShadeLen*\uCy}) -- cycle;
  \fill[orange!45, opacity=0.4] (axis cs:\cOneMin,\cTwoMin) rectangle (axis cs:\cOneMax,\cTwoMax);
\end{scope}

\begin{scope}
  \clip (axis cs:0,0) -- (axis cs:{\ShadeLen*\uCx},{\ShadeLen*\uCy})
        -- (axis cs:{\ShadeLen*\uDx},{\ShadeLen*\uDy}) -- cycle;
  \fill[cyan!50, opacity=0.4] (axis cs:\cOneMin,\cTwoMin) rectangle (axis cs:\cOneMax,\cTwoMax);
\end{scope}

\begin{scope}
  \clip (axis cs:0,0) -- (axis cs:{\ShadeLen*\uDx},{\ShadeLen*\uDy})
        -- (axis cs:{\ShadeLen*\uAx},{\ShadeLen*\uAy}) -- cycle;
  \fill[teal!40, opacity=0.4] (axis cs:\cOneMin,\cTwoMin) rectangle (axis cs:\cOneMax,\cTwoMax);
\end{scope}

\addplot[black!60, very thin] coordinates {(0,\cTwoMin) (0,\cTwoMax)};
\addplot[black!60, very thin] coordinates {(\cOneMin,0) (\cOneMax,0)};
\addplot[mark=+,only marks, mark size=4pt, line width=0.5pt] coordinates {(0,0)};

\plotSafeRay{\uDAxRaw}{\uDAyRaw}  
\plotSafeRay{\uABxRaw}{\uAByRaw}  
\plotSafeRay{\uBCxRaw}{\uBCyRaw}  
\plotSafeRay{\uCDxRaw}{\uCDyRaw}  

\addplot[magenta, very thick, -{Stealth[length=3mm,width=2mm]}]
  coordinates {(0,0) (\RayLenCost*\uEOneTwoX,\RayLenCost*\uEOneTwoY)};
\addplot[magenta, very thick, -{Stealth[length=3mm,width=2mm]}]
  coordinates {(0,0) (\RayLenCost*\uETwoThreeX,\RayLenCost*\uETwoThreeY)};
\addplot[magenta, very thick, -{Stealth[length=3mm,width=2mm]}]
  coordinates {(0,0) (\RayLenCost*\uEThreeFourX,\RayLenCost*\uEThreeFourY)};
\addplot[magenta, very thick, -{Stealth[length=3mm,width=2mm]}]
  coordinates {(0,0) (\RayLenCost*\uEFourOneX,\RayLenCost*\uEFourOneY)};

\pgfmathsetmacro{\edgeLabelPad}{7.5}
\node[edgelabel] at (axis cs:{(\RayLenCost+\edgeLabelPad)*\uEOneTwoX},
                             {(\RayLenCost+\edgeLabelPad)*\uEOneTwoY}) {\ELabelTwelve};
\node[edgelabel] at (axis cs:{(\RayLenCost+\edgeLabelPad)*\uETwoThreeX},
                             {(\RayLenCost+\edgeLabelPad)*\uETwoThreeY}) {\ELabelTwentyThree};
\node[edgelabel] at (axis cs:{(\RayLenCost+\edgeLabelPad)*\uEThreeFourX},
                             {(\RayLenCost+\edgeLabelPad)*\uEThreeFourY}) {\ELabelThirtyFour};
\node[edgelabel] at (axis cs:{(\RayLenCost+\edgeLabelPad)*\uEFourOneX},
                             {(\RayLenCost+\edgeLabelPad)*\uEFourOneY}) {\ELabelFortyOne};

\pgfmathsetmacro{\regionLabelScale}{0.15}
\newcommand{\midUnit}[4]{%
  \pgfmathsetmacro{\mx}{#1 + #3}%
  \pgfmathsetmacro{\my}{#2 + #4}%
  \pgfmathsetmacro{\mm}{sqrt(\mx*\mx + \my*\my)}%
  \pgfmathsetmacro{\mux}{\mx/\mm}%
  \pgfmathsetmacro{\muy}{\my/\mm}%
}

\midUnit{\uAx}{\uAy}{\uBx}{\uBy} \let\uxAB\mux \let\uyAB\muy
\midUnit{\uBx}{\uBy}{\uCx}{\uCy} \let\uxBC\mux \let\uyBC\muy
\midUnit{\uCx}{\uCy}{\uDx}{\uDy} \let\uxCD\mux \let\uyCD\muy
\midUnit{\uDx}{\uDy}{\uAx}{\uAy} \let\uxDA\mux \let\uyDA\muy

\pgfmathsetmacro{\regionLabelR}{2.8}
\pgfmathsetmacro{\regionLabelRAdjX}{\regionLabelR * \regionLabelScale * \dxC}
\pgfmathsetmacro{\regionLabelRAdjY}{\regionLabelR * \regionLabelScale * \dyC}

\node[vertexlabel] at (axis cs:{\regionLabelRAdjX*\uxAB},{\regionLabelRAdjY*\uyAB}) {\RLabelAB};
\node[vertexlabel] at (axis cs:{\regionLabelRAdjX*\uxBC},{\regionLabelRAdjY*\uyBC}) {\RLabelBC};
\node[vertexlabel] at (axis cs:{\regionLabelRAdjX*\uxCD},{\regionLabelRAdjY*\uyCD}) {\RLabelCD};
\node[vertexlabel] at (axis cs:{\regionLabelRAdjX*\uxDA},{\regionLabelRAdjY*\uyDA}) {\RLabelDA};

\addplot[blue, ultra thick, -{Latex[length=2mm,width=2mm]}] 
coordinates {(0,0) (\cOneVec,\cTwoVec)} 
node[pos=1, anchor=south west, font=\normalsize, xshift=-3.5mm, yshift=-3.5mm] {$r$};
    
\end{axis}

\end{tikzpicture}